\documentclass[aps,superscriptaddress,showpacs,preprintnumbers,twocolumn]{revtex4-1}
\usepackage{amsmath}
\usepackage{graphicx}
\usepackage{draftcopy}
\begin{document}

\preprint{LA-UR-09-07764}
\title{Nonparametric Dark Energy Reconstruction from Supernova Data} 

\author{Tracy Holsclaw}
\affiliation{Department of Applied Mathematics and Statistics, 
 University of California, Santa Cruz, CA 95064}
\author{Ujjaini Alam}
\affiliation{ISR-1, MS D466, Los Alamos National Laboratory, Los
Alamos, NM 87545}
\author{Bruno Sans\'o}
\affiliation{Department of Applied Mathematics and Statistics, 
 University of California, Santa Cruz, CA 95064}
\author{Herbert Lee}
\affiliation{Department of Applied Mathematics and Statistics, 
 University of California, Santa Cruz, CA 95064}
\author{Katrin Heitmann}
\affiliation{ISR-1, MS D466, Los Alamos National Laboratory, Los
Alamos, NM 87545}
\author{Salman Habib}
\affiliation{T-2, MS B285, Los Alamos National Laboratory, Los
Alamos, NM 87545}
\author{David Higdon}
\affiliation{CCS-6, MS F600, Los
Alamos National Laboratory, Los Alamos, NM 87545}

\date{\today}

\begin{abstract}
  Understanding the origin of the accelerated expansion of the
  Universe poses one of the greatest challenges in physics today.
  Lacking a compelling fundamental theory to test, observational
  efforts are targeted at a better characterization of the underlying cause.
  If a new form of mass-energy, dark energy, is driving the
  acceleration, the redshift evolution of the equation of state parameter $w(z)$
  will hold essential clues as to its origin. To best exploit data
  from observations it is necessary to develop a robust and accurate
  reconstruction approach, with controlled errors, for $w(z)$. We
  introduce a new, nonparametric method for solving the associated
  statistical inverse problem based on Gaussian Process modeling and
  Markov chain Monte Carlo sampling. Applying this method to recent
  supernova measurements, we reconstruct the continuous history of $w$
  out to redshift $z=1.5$.
\end{abstract}

\pacs{98.80.-k, 02.50.-r}

\maketitle

Little more than a decade has passed after supernova observations
first found evidence for the accelerated expansion of the Universe
~\cite{riessperl}. Since confirmed by different probes, this
remarkable discovery has been hailed as the harbinger of a revolution
in fundamental physics and cosmology. Cosmic acceleration demands
completely new physics -- it challenges basic notions of quantum
theory, general relativity, and assumptions regarding the fundamental
make-up of matter. Currently, the two most popular explanations are a
dark energy, usually modeled by a scalar field, or a modification of
general relativity on cosmic length scales. In the absence of a
compelling candidate theory to explain the observations, the target of
current and future cosmological missions is to characterize the
underlying cause for the accelerated expansion. In the case of dark
energy, the aim is to constrain the equation of state parameter
$w=p/\rho$ and its possible evolution. A deviation from $w=const.$
would provide clues pointing to the origin of the accelerated
expansion. (Currently, observations are consistent with a cosmological
constant, $w=-1$, at the 10\% level~\cite{hicken09}.)

In order to extract useful information from cosmological data, a
reliable and robust reconstruction method for $w(z)$ is crucial. Here,
we introduce a nonparametric technique based on Gaussian Process (GP)
modeling and Markov chain Monte Carlo (MCMC) sampling, and apply it to
supernova data. GPs extend the multivariate Gaussian distribution to
function spaces, with inference taking place in the space of
functions. The defining property of a GP is that the vector that
corresponds to the process at any finite collection of points follows
a multivariate Gaussian distribution. Gaussian processes are elements
of an infinite dimensional space, and can be used as the basis for a
nonparametric reconstruction method. GPs are characterized by mean and
covariance functions, defined by a small number of
hyperparameters~\cite{gprefs}. The covariance function controls
aspects such as roughness of the candidate functions and the length
scales on which they can change; aside from this, their shapes are
arbitrary. Bayesian estimation simultaneously evaluates the GP
hyperparameters (so-called to prevent confusion with the parameters
that define a parametric method) together with quantities of physical
interest.

For supernovae, the reconstruction task can be summarized as
follows.  The data is given in the form of the distance modulus
$\mu_B(z)$ defined as:
\begin{equation}\label{mu}
\mu_B(z)=m_B-M_B=5\log_{10}\left(\frac{d_L (z)}{1 {\rm Mpc}}\right)+25,
\end{equation} 
where $m_B$ and $M_B$ are the apparent and absolute magnitudes.
The luminosity distance $d_L(z)$ is connected to the Hubble expansion
rate $H(z)$, and thus to $w(z)$, via: 
\begin{eqnarray}\label{dl}
d_L(z)&=&(1+z)\frac{c}{H_0}\int_0^z\frac{ds}{h(s)}\\
&=&(1+z)\frac{c}{H_0}
\int^z_0 ds\left[\Omega_m(1+s)^{3}\right.\nonumber\\
&&\left.+(1-\Omega_m)(1+s)^3
\exp\left(3\int_0^{s}\frac{w(u)}{1+u}du\right)\right]^{-\frac{1}{2}},\nonumber
\end{eqnarray}
where $h(z)=H(z)/H_0$. Note that supernovae cannot be used to
determine $H_0$ in the absence of an independent distance measurement.
Thus it remains an unknown and can be absorbed in a redefinition of
the absolute magnitude: ${\cal M_B}=M_B-5\log_{10}(H_0)+25$. The $H_0$
used to obtain ${\cal M_B}$ is not the physical value, but a certain
fixed constant assumed in the observational analysis. For the dataset
analyzed below~\cite{hicken09}, $H_0=65$ km/s/Mpc. Since we will work
with $\mu_B$ and not with $m_B$ directly, the numerical value for
${\cal M_B}$ does not enter in our analysis. We allow for a free
constant, $\Delta_\mu$ in Eqn.~(\ref{mu}), accounting for the
uncertainty in the absolute calibration of the data, and for which we
choose a uniform prior between $[-0.5,0.5]$. We assume spatial
flatness as an ``inflation prior''; strong constraints on spatial
flatness exist from combining cosmic microwave background (CMB) and
baryon acoustic oscillation (BAO) measurements~\cite{komatsu}. The
prior on $\Omega_m$ is also informed by the 7-year WMAP
analysis~\cite{komatsu} for a $w$CDM model combining CMB, BAO, and
$H_0$ measurements. Since our assumptions on $w$ are less strict than
$w=const.$ we broaden the nominal range by a factor of two, leading to
a Gaussian prior with mean 0.27 and standard deviation of 0.04.

Reconstruction is a classic statistical inverse problem for the
nonlinear (smoothing) operator of Eqn.~(\ref{dl}), where one solves
for the function $w(z)$ and for the parameter $\Omega_m$, given a
finite set of noisy data for $d_L(z)$. Different strategies may be
followed to arrive at a tractable formulation. (i) Assume a
parameterized form for $w(z)$ and estimate the associated parameters.
This approach is the most commonly used currently; the parametric
forms either assume $w=const.$ or allow for a fixed redshift variation
such as $w=w_0-w_1z/(1+z)$, where $w_0$ and $w_1$ are
constants~\cite{linder03}. (ii) Pick a simple local basis
representation for $w(z)$ (bins, wavelets), and estimate the
associated coefficients (effectively a piecewise constant
description), using Principal Component Analysis (PCA) if needed, to
work with eigenmodes defined as linear combinations of
bins~\cite{huterer04,hojjati09}. (iii) Follow a procedure similar to
(ii) -- without PCA -- but actually use (filtered) numerical
derivatives to estimate $w(z)$~\cite{daly03}. (iv) Use a distribution
over (random) functions that can represent $w(z)$ and estimate the
statistical properties thereof, given observed data. Methods (i),
(ii), and (iv) can all be carried out using a Bayesian viewpoint and
exploring posteriors by MCMC methods, whereas (iii) -- as presented in
the literature -- represents a different class of approach to the
inverse problem. Taking numerical derivatives is generally a difficult
task and a corresponding error theory seems hard to develop. Approach
(i) can encounter difficulties if $w(z)$ has a nontrivial evolution.
The finite parameterization and the specific functional form assumed
can bias results for the temporal behavior of $w(z)$~\cite{simpson06}.
Methods (ii) and (iv) apply different philosophies -- (ii) applies a
local view of the reconstruction ($z$ bins), whereas (iv) attempts to
sample the posterior continuously in $z$. In a mild sense, the choice
of a piecewise continuous representation in (ii) -- of which,
$w=const.$ is just the one-bin limit -- can force an unphysical view
of $w(z)$, since the actual $w(z)$ is not piecewise constant. In
contrast, method (iv) while fully nonparametric, is potentially more
general and flexible compared to the other methods.

Our new GP modeling-based approach is a realization of method (iv). It
enables the identification of nontrivial redshift dependences in $w(z)$
reliably, if they exist (Ref.~\cite{holsclaw09} shows examples based
on simulated data). The central idea is to assign prior probabilities to
classes of functions via GPs and to take advantage of the particular
integral structure of Eqn.~(\ref{dl}), again using GPs. Employing a
Bayesian approach to explore posterior distributions over the
functions via MCMC we not only obtain a family of continuous
realizations for $w(z)$ but at the same time optimize the GP model
hyperparameters, informed by the actual data, comprehensively
propagating all estimation uncertainties.

One may wonder whether a general nonparametric reconstruction must
involve taking a second derivative of the data in some way.
 This is true only in
a formal sense -- the approach described here does not involve any
derivatives. Instead, we invert an integral equation, ill-posed
because the operator to be treated is a complicated smoothing operator
involving two integrals. To make the problem well-behaved we make mild
continuity assumptions about $w(z)$ which are justified if the origin
of dark energy is to be described by a reasonable physical model.

\begin{figure}
\centerline{
  \includegraphics[width=1.6in]{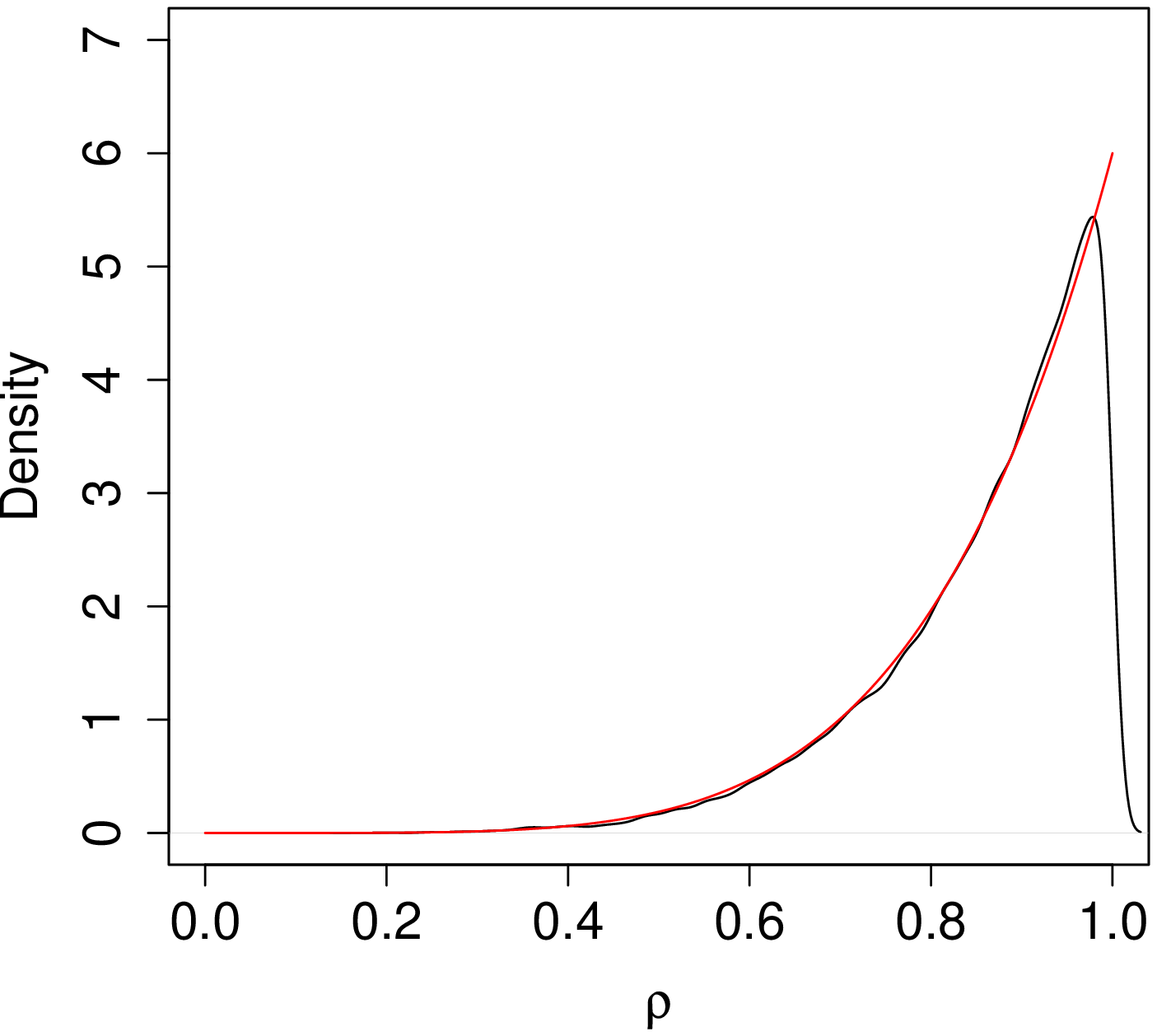}
  \includegraphics[width=1.6in]{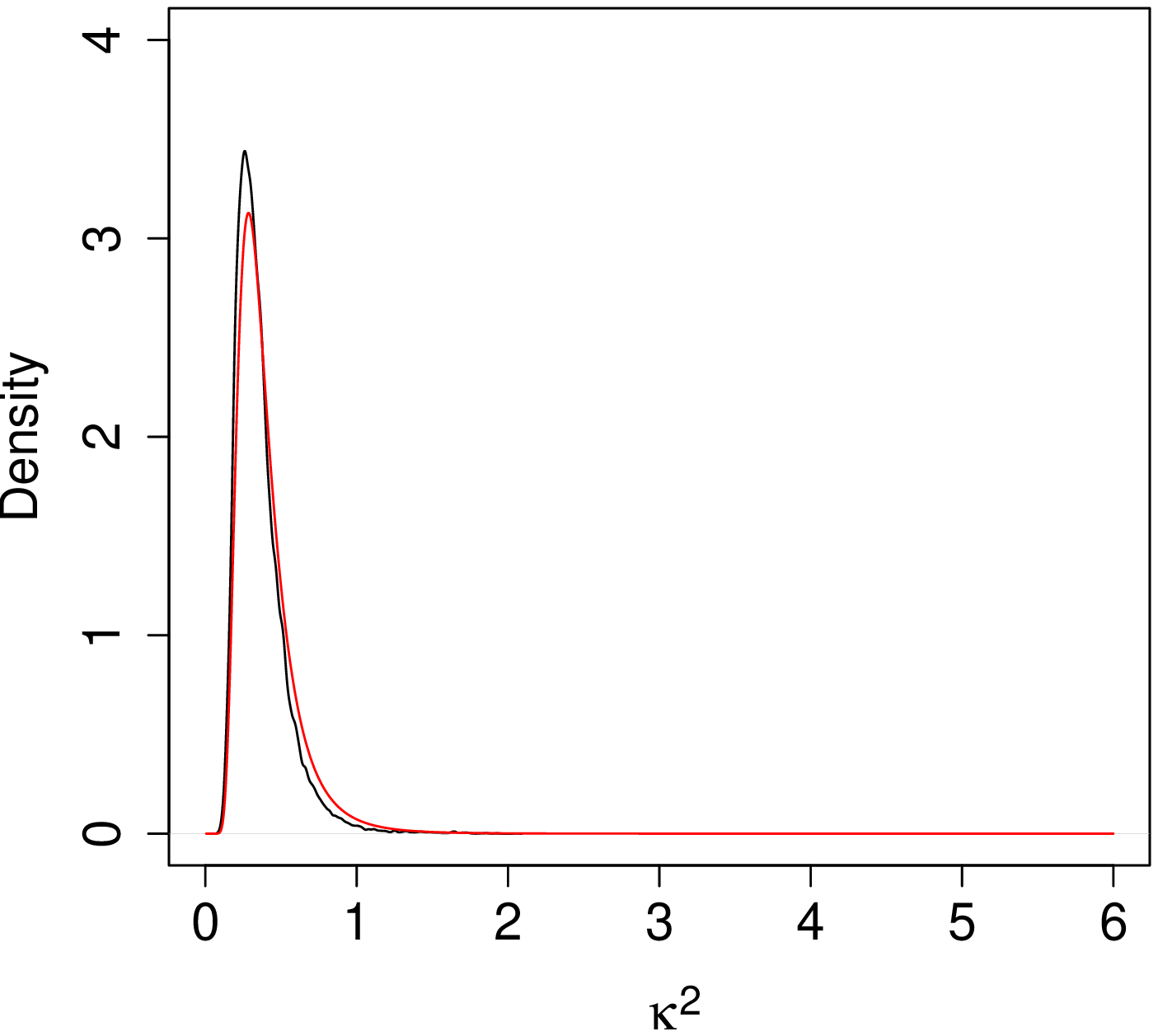}}
  \includegraphics[width=1.6in]{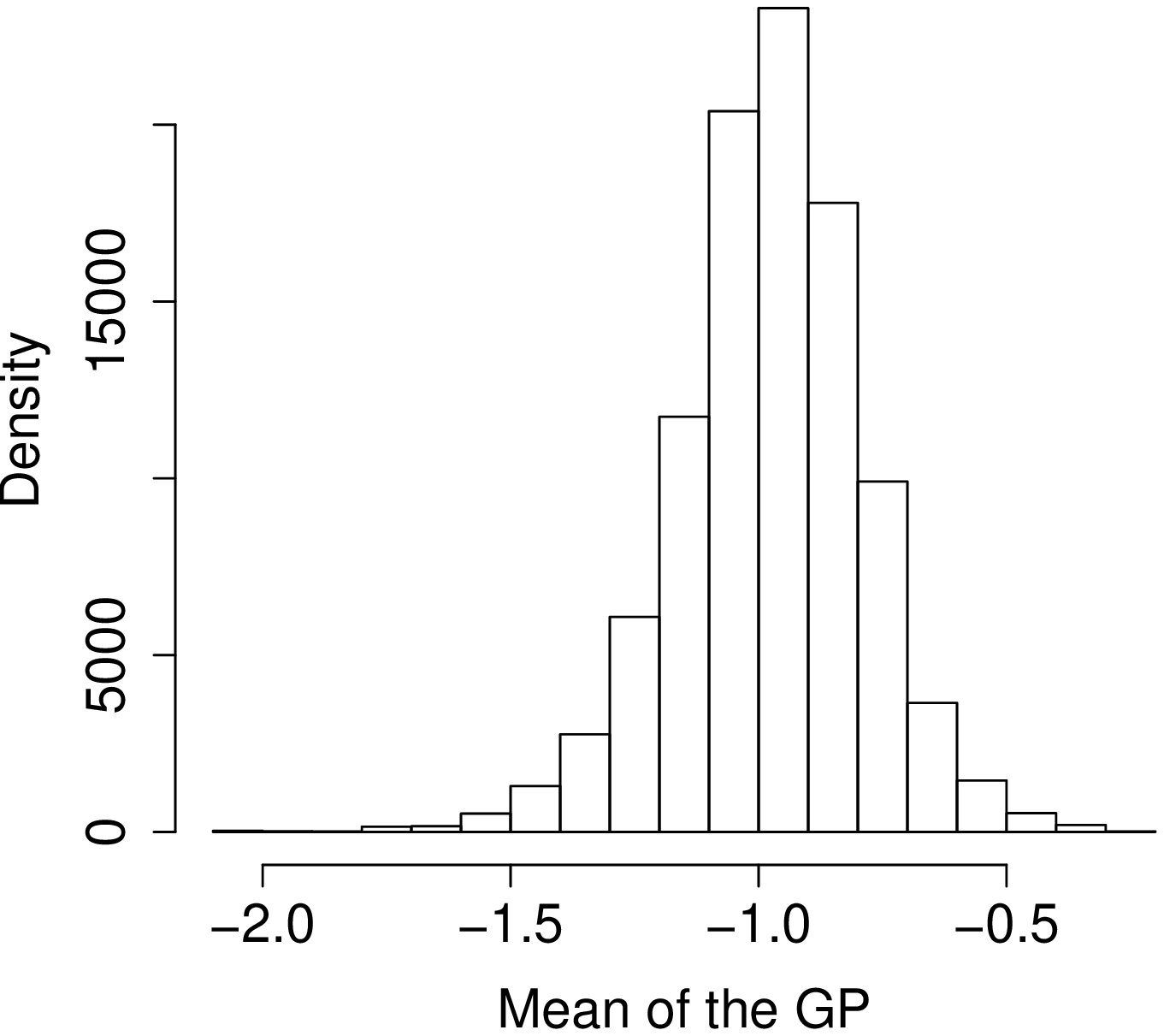}
   \includegraphics[width=1.6in]{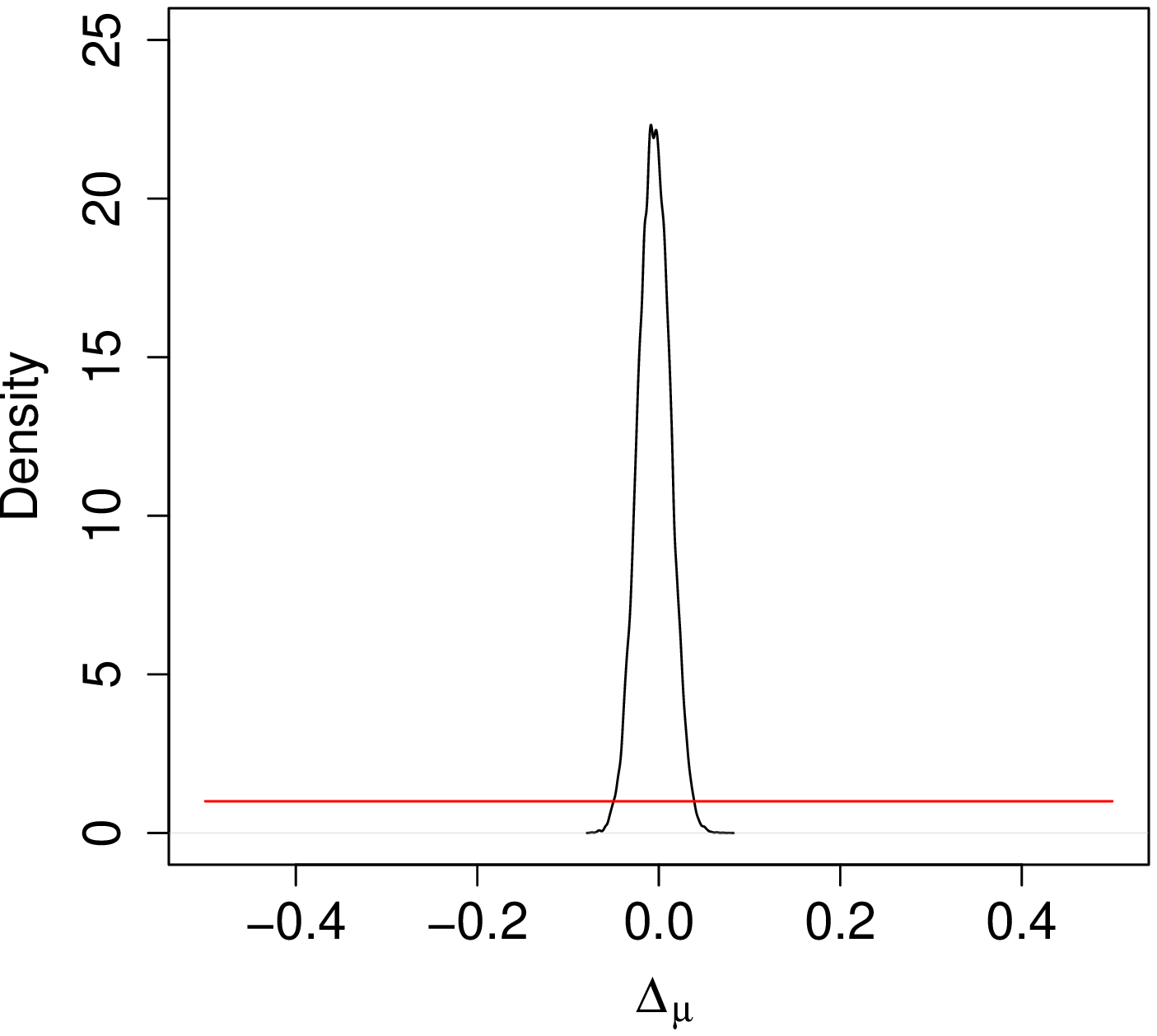}
  \caption{\label{priors} Priors (red lines) and posteriors (black
  lines) for the GP hyperparameters $\rho$ and
  $\kappa^2$. The lower left panel shows the distribution of the GP
  mean. The lower right panel shows the results for $\Delta_\mu$. 
  The posteriors for different $\alpha$ choices are very similar, and 
  we show only the results for $\alpha \simeq 2$ here.}
\end{figure}

As previously noted, a GP model assumes that $w(z_1),..., w(z_n)$, for
any set of redshifts $z_1,...,z_n$, follow a multivariate Gaussian
distribution specified by mean and covariance functions~\cite{gprefs}.
Here we use a mean of negative one as the prior and exponential family
covariance functions written as ($\rho$ being a numerical constant):
\begin{equation}
K(z,z')=\kappa^2\rho^{|z-z'|^\alpha}.  \nonumber
\end{equation}
The hyperparameters $\rho \in (0,1)$ and $\kappa$, and the parameters
defining the likelihood, are determined by the data. The value of
$\alpha\in(0, 2]$ influences the smoothness of the GP realizations:
for $\alpha=2$, the realizations are smooth with infinitely many
derivatives (this covariance function corresponds to using an infinite
number of Gaussian basis functions), while $\alpha=1$ leads to rougher
realizations suited to modeling continuous
non-(mean-squared)-differentiable functions. We use both $\alpha$
values in our analysis, the results being very similar. The form of
the covariance function is unrelated to the shape of the GP sampling
functions, as determined by the data, so no particular behavior is
assumed for $w(z)$. There is no loss of generality in fixing the
(statistical ensemble) GP mean, as any variation imposed by the data
appears in the covariance function. Fixing the mean has the advantage
of improving the stability of the MCMC (we explored other means and
found very similar results). We stress that even though the mean is
fixed, each GP realization will actually have a different mean with a
spread controlled by $\kappa$ as shown in Figure~\ref{priors}. The
constant $\rho$ has a prior of $Beta(6,1)$ and $\kappa^2$ has a vague
Inverse Gamma prior $IG(6,2)$. The probability distribution of the
$Beta$ prior is given by
$f(x;\alpha,\beta)=\Gamma(\alpha+\beta)x^{\alpha-1}(1-x)^{\beta-1}/
[\Gamma(\alpha)\Gamma(\beta)]$ and for the $IG$ prior by
$f(x;\alpha,\beta)=\beta^{\alpha}x^{-\alpha-1}
\Gamma(\alpha)^{-1}\exp(-\beta/x)$, with $x>0$. We show the priors for
$\rho$ and $\kappa^2$ and their posterior distribution in
Figure~\ref{priors}.

Following the notation of Eqn.~(\ref{dl}) we set up the following
GP for $w$:
\begin{equation}
w(u)\sim {\rm GP}(-1,K(u,u')). 
\nonumber
\end{equation}
 Recall that we have to integrate over $w(u)$ (Eqn.~\ref{dl}):
\begin{equation}
\label{int}
y(s)=\int_0^s\frac{w(u)}{1+u}du.
\end{equation}
We use the fact that the integral of a GP is also a GP with mean and
covariance dependent on the original GP~\cite{gprefs}. Using that
result, we set up a second GP for $y(s)$: 
\begin{equation}
y(s)\sim {\rm GP}\left(-\ln (1+s),\kappa^2
\int_0^s\int_0^{s'}\frac{\rho^{|u-u'|^\alpha}dudu'}{(1+u)(1+u')}\right) .\nonumber
\end{equation}
The mean value for this GP is simply obtained by solving the integral
in Eqn.~(\ref{int}) for the mean value of the GP for $w(u)$, here
taken to be negative one. As mentioned earlier, even though the
ensemble mean is fixed to a constant value at any $z$, each GP
realization is not mean-zero (over $z$). We show the distribution of
the mean for the different realizations in Figure~\ref{priors}
demonstrating a wide spread between -2 and -0.6. In addition, we used
simulated data with a time-varying equation of state to check that the
choice of the mean does not bias the results.

\begin{figure}
\includegraphics[width=2.3in]{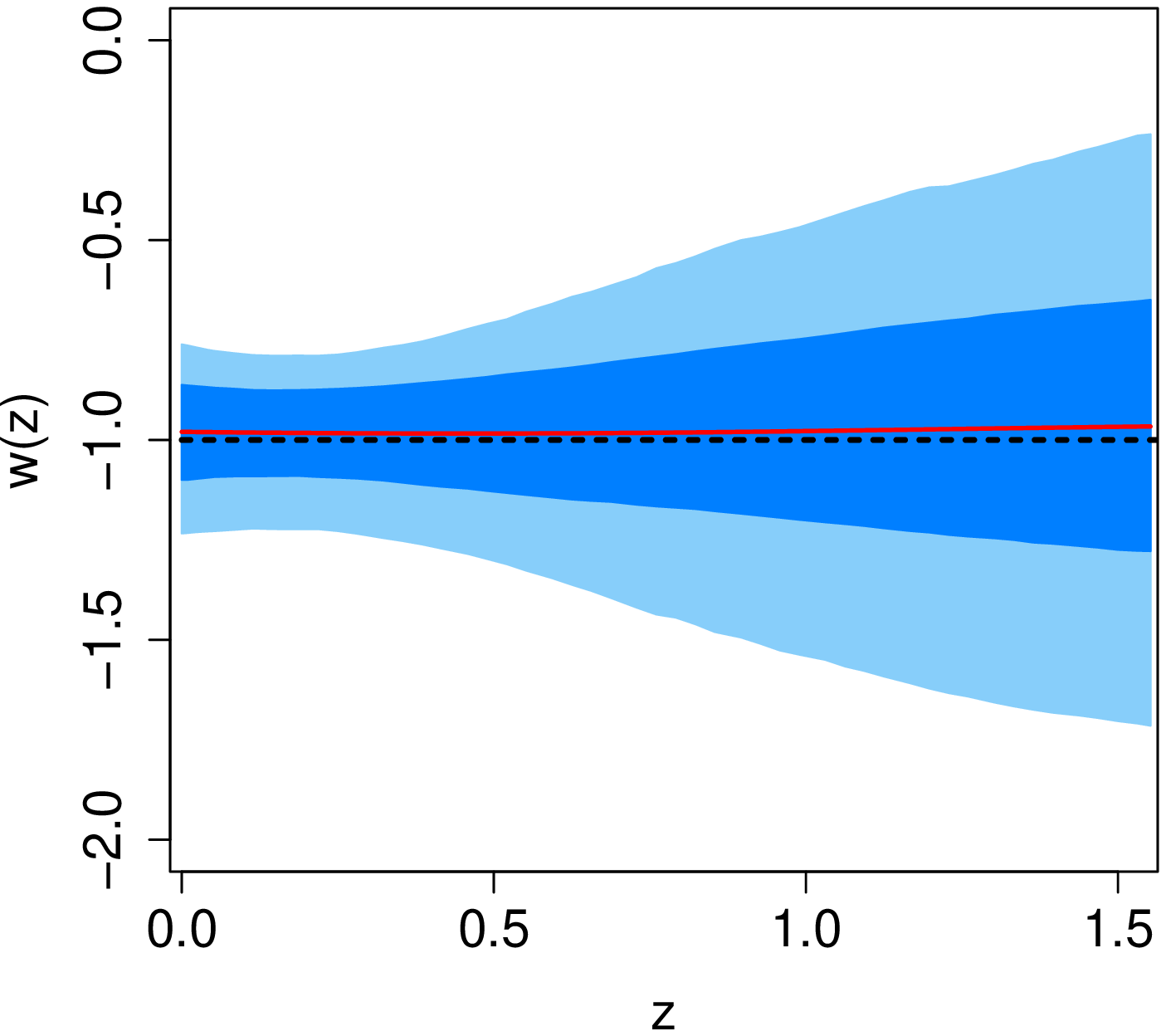}
\includegraphics[width=2.3in]{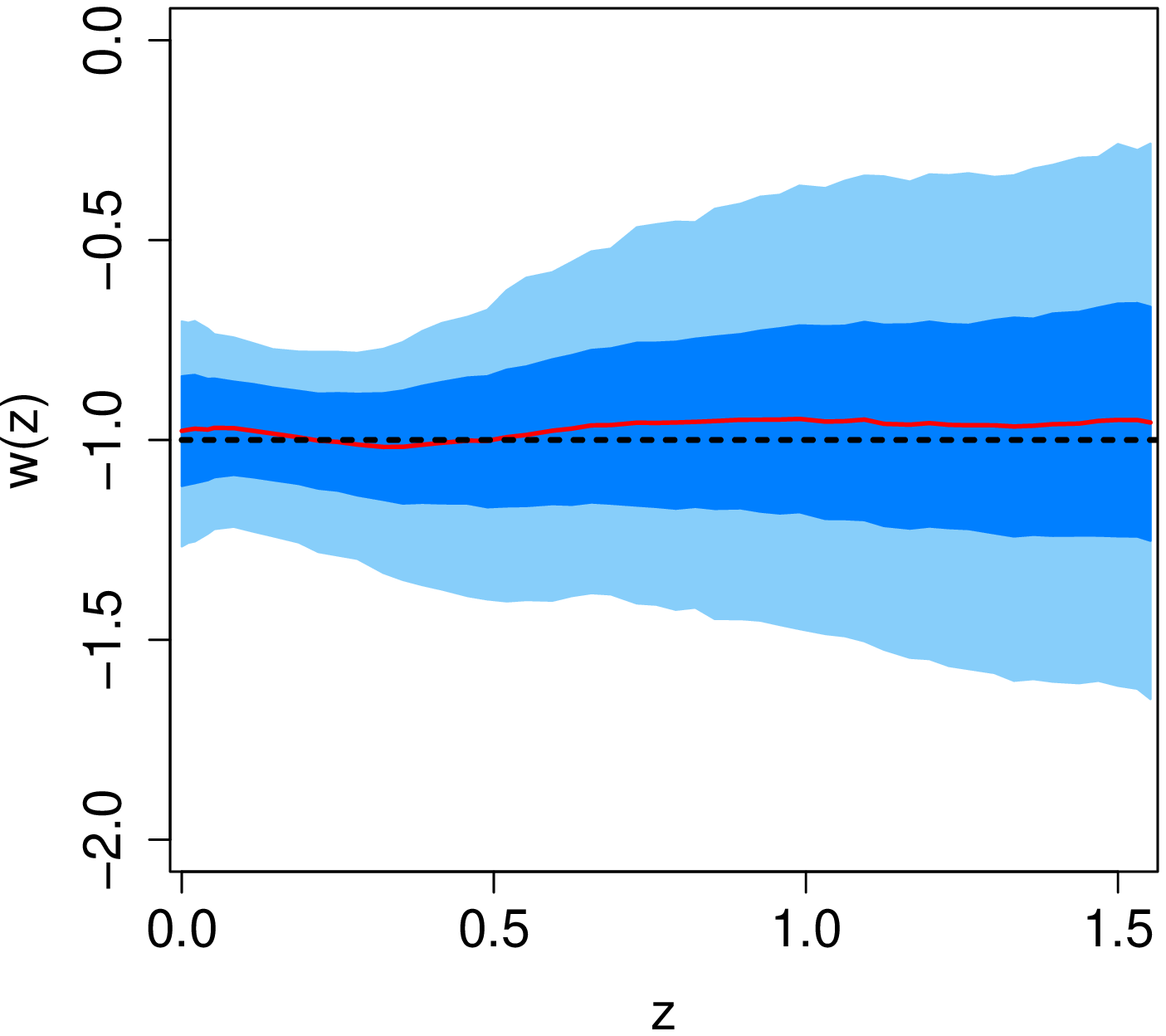}
\caption{\label{wz_gp}Nonparametric reconstruction of $w(z)$ based on
  GP modeling combined with MCMC. The upper panel uses a Gaussian
  covariance function, the bottom panel, an exponential covariance
  function. Both results are very close and in agreement with a
  cosmological constant (black dashed line). The dark blue shaded
  region indicates the 68\% confidence level, while the light blue
  region extends it to 95\%.} 
\end{figure}

We can now construct a joint GP for $y(s)$ and $w(u)$:
\begin{equation}
\left[\begin{array}{c}y(s)\\w(u)\end{array}\right]\sim{\rm
MVN}\left[\left[
\begin{array}{c}-\ln(1+s)\\-1\end{array}\right]
\left[\begin{array}{cc}\Sigma_{11} & \Sigma_{12}\\\Sigma_{21}&
\Sigma_{22}\end{array}\right]\right],\nonumber
\end{equation}
a multi-variate normal (MVN) distribution with
\begin{eqnarray}
\Sigma_{11}&=&\kappa^2\int_0^s\int_0^{s'}\frac{\rho^{|u-u'|^\alpha}dudu'}{(1+u)(1+u')}, 
\label{sig11}\\ 
\Sigma_{22}&=&\kappa^2\rho^{|s-s'|^\alpha},\\
\Sigma_{12}&=&\kappa^2\int_0^s\frac{\rho^{|u-u'|^\alpha}du}{(1+u)}.
\end{eqnarray}
The mean for $y(s)$ given $w(u)$ can be found through the following
relation:
\begin{equation}
\langle y(s)|w(u)\rangle=-\ln(1+s)+\Sigma_{12}\Sigma_{22}^{-1}\left[w(u)-(-1)\right]. \nonumber
\end{equation}
Now only the outer integral is left to be solved for in
Eqn.~(\ref{dl}), and this can be computed by standard numerical
methods. (Note that the computationally expensive double integral for
$\Sigma_{11}$ as defined in Eqn.~(\ref{sig11}) does not need to be
performed.) The GP prior can now be combined with a likelihood
function to obtain a posterior that can be sampled by MCMC methods.
Details of our GP-based MCMC implementation are provided in the
supplementary material~\cite{supp}.

For our specific analysis we focus on a recent composite supernova
dataset, provided by Hicken et al.~\cite{hicken09}. This dataset
combines the so-called Union dataset~\cite{union} with new
measurements of low redshift supernovae to form the Constitution set.
The dataset has been analyzed in Ref.~\cite{hicken09} using different
light curve fitters for the supernova light curves; our analysis uses
results from the SALT fitter (Table 1 in Ref.~\cite{hicken09}, which 
includes estimates for the error for the distance modulus $\mu_B$ --
the tables contain what is referred to in Ref.~\cite{hicken09} as the
``minimal cut''). 

Our final results for $w(z)$ are shown in Figure~\ref{wz_gp}. The
upper panel shows the results from a GP model with a Gaussian
covariance function ($\alpha\simeq 2$) while the results in the lower
panel are based on an exponential covariance function ($\alpha=1$).
The results are very similar, the Gaussian covariance function leading
to a slightly smoother prediction. The mean value of $w(z)$ is very
close to -1 at redshifts close to zero and rises slightly at redshift
$z=1.5$. Within our estimated errors, the results are consistent with
a cosmological constant $w=-1$. Note, however, that realizations of
$w(z)$ with nontrivial $z$-dependence are not excluded; as
observations improve the allowed range of variability will be further
constrained. In Ref.~\cite{hicken09} a combined analysis of supernova
data and baryon acoustic oscillation measurements is carried out.
Assuming $w=const.$, the SALT-based dataset yields
$w=-0.987^{+0.066}_{-0.068}$ consistent with our findings.

To summarize, we have presented a new, nonparametric reconstruction
technique for the dark energy equation of state and applied it to
current supernova observations. The GP-based method can be used to
determine the most probable behavior of $w(z)$ and to infer how likely
a target trajectory is given the current data. Thus it can be used to
accept or reject classes of $w(z)$ models. The method allows adjusting
of smoothness assumptions regarding $w(z)$; priors on the GP
hyperparameters control the allowed arbitrariness (e.g., degree of
differentiability). Robustness of the results obtained can be checked
by varying these priors. Our results for $w(z)$ are consistent with a
cosmological constant, with no evidence for a systematic mean
evolution in $w$ with redshift, although variations within our error
limits cannot be ruled out. We have carried out careful tests to
ensure that our choices of priors and hyperparameters do not alter the
results. Our method possesses several advantages: it avoids artificial
biases due to restricted parametric assumptions for $w(z)$, it does
not lose information about the data by smoothing it, and it does not
introduce arbitrariness (and lack of error control) in reconstruction
by representing the data using a certain number of bins, or cutting
off information by using a restricted set of basis functions to
represent the data. The technique can be easily extended to fold in
data from CMB and BAO observations; work in this direction is
currently in progress. The GP-based MCMC procedure can be integrated
within supernova analysis frameworks, e.g., SNANA~\cite{snana} as a
cosmology fitter, following the general methodology presented in
Ref.~\cite{habib07}.

We thank the LANL/UCSC Institute for Scalable Scientific Data
Management for supporting this work. Partial support by the DOE under
contract W-7405-ENG-36 is also acknowledged. UA, SH, KH, and DH
acknowledge support from the LDRD program at LANL; KH acknowledges
support from the NASA Theory program. SH and KH thank the Aspen Center
for Physics, where part of this work was carried out. We are indebted
to Andy Albrecht, Josh Frieman, Adrian Pope, Martin White, and Michael
Wood-Vasey for insightful discussions.

\end{document}